\documentclass{appolb}

\usepackage{amssymb,amsmath,color,hyperref,graphicx,multirow,tabularx,physics}
\usepackage{davitex}
\usepackage{lineno}

\newcommand{\Pbare}{P^{\text{bare}}}

\newcommand{\freg}{f_{\rm reg}}

\usepackage{hyperref}
\hypersetup{
  colorlinks=true,
  linkcolor=black,
  citecolor=black,
  urlcolor=blue,
}

\usepackage[
backend=bibtex,
sorting=none,
style=numeric,
url=false,
giveninits
]{biblatex}
\AtEveryBibitem{%
  \clearfield{month}%
}
\DeclareFieldFormat{pages}{#1}
\DeclareBibliographyDriver{article}{%
  \usebibmacro{begentry}%
  \usebibmacro{author/translator+others}%
  \addcomma\space
  \usebibmacro{journal}%
  \addcomma
  \usebibmacro{volume+number+eid}%
  \usebibmacro{note+pages}%
  \usebibmacro{issue+date}%
  \usebibmacro{finentry}}
\begin{document}

\title{Sensitivity of the Polyakov loop to\\ chiral symmetry 
restoration\thanks{Presented at workshop on Criticality in QCD and 
the Hadron Resonance Gas; 29-31 July 2020, Wroclaw, Poland.}%
\headtitle{Polyakov loop in the chiral limit}
}

\author{D. A. Clarke\thanks{Speaker.}, O. Kaczmarek, Anirban Lahiri, 
Mugdha Sarkar
\headauthor{Clarke et al.}
\address{Fakult\"at f\"ur Physik, Universit\"at Bielefeld, D-33615
Bielefeld, Germany}
}

\maketitle

\begin{abstract}
In the heavy, static quark mass regime of QCD, the Polyakov loop is well known
to be an order parameter of the deconfinement phase transition; however, the
sensitivity of the Polyakov loop to the deconfinement of light, dynamical
quarks is less clear. On the other hand, from the perspective of an effective
Lagrangian written in the vicinity of the chiral transition, the Polyakov
loop is an energy-like operator and should hence scale as any energy-like
operator would. We show here that the Polyakov loop and 
heavy-quark free energy are sensitive to the chiral transition, i.e.
their scaling is consistent with energy-like observables in 3-$d$ $\O(N)$ 
universality classes.
\end{abstract}


%%%%%%%%%%%%%%%%%%%%%%%%%%%%%%%%%%%%%%%%%%%%%%%%%%%%%%%%%%%%%%%%%%%%%%%%%%%%%%
%  Introduction
%%%%%%%%%%%%%%%%%%%%%%%%%%%%%%%%%%%%%%%%%%%%%%%%%%%%%%%%%%%%%%%%%%%%%%%%%%%%%%
\section{Introduction}\label{sec:intro}

The QCD Lagrangian with two mass-degenerate light quarks $m_l$ possesses an 
exact global $\mathbb{Z}_3$ symmetry in the quenched limit, i.e. in the
infinite quark mass limit. At low temperature
where the symmetry is respected, quarks are confined
into bound states; above a critical temperature $T_d$, 
the $\mathbb{Z}_3$ symmetry breaks spontaneously, and quarks become 
deconfined. The average Polyakov loop\footnote{Note that $\ev{\Im P}=0$ 
at all values of $T$ and all quark masses as the Euclidean QCD action 
is invariant under $U_{\mu}(\vec{x},\tau)\to U_{\mu}^\dagger (\vec{x},\tau)$.} 
$\ev{P}=\ev{\Re P}$ serves as a natural
order parameter for this deconfinement transition.
At the other end of the spectrum, where $m_l=0$, the Lagrangian has an
$\SU(2)_L\times\SU(2)_R$ symmetry that spontaneously breaks below a
critical temperature $T_c$. In this limit, the light quark chiral
condensate $\ev{\bar{\psi}\psi}$ serves as an appropriate order parameter.

At physical $m_l$ both symmetries are broken explicitly, and neither
$\ev{P}$ nor $\ev{\bar{\psi}\psi}$ is zero. Still, these quantities
will change substantially as a function of temperature, and
the inflection points of these changes have been used to define pseudo-critical
temperatures. In the quenched limit \cite{kogut_deconfinement_1983} and also at 
larger-than-physical quark mass values \cite{cheng_qcd_2008}, 
inflection points in the temperature dependence for $\ev{P}$ and
$\ev{\bar{\psi}\psi}$ have been found at similar
temperatures. This seeming coincidence of inflection points
is often taken as evidence for the coincidence of chiral and 
deconfinement transitions.
However, studies with (almost) physical light quark masses and, 
in particular, studies with improved fermion actions, performed closer to 
the continuum limit, in general show that the QCD transition is a smooth 
crossover, and no coincidence of inflection points is found
\cite{aoki_qcd_2009,Bazavov:2016uvm,clarke_polyakov_2020}, challenging
this line of evidence for simultaneous chiral and deconfinement transitions.

This way of thinking interprets the rapid change in $\ev{P}$ as a remnant
of $m_l=\infty$ physics. But in the chiral limit, there is no obvious
symmetry whose breaking can be related to deconfinement; in that sense,
there is no {\it a priori} reason to interpret $\ev{P}$ in the
chiral limit in this way.
Another possibility is that the behavior of $\ev{P}$ is instead 
sensitive to the chiral transition in this regime.
Under this assumption, $\ev{P}$ should inherit its behavior
from the chiral transition as an energy-like operator. 
%This is because the Polyakov loop is trivially invariant under chiral 
%rotations, so if one were to write an effective QCD Lagrangian near the 
%chiral limit, the Polyakov loop should be hidden in the symmetry-respecting
%energy-like term rather than the symmetry-breaking magnetization-like term.

In this study we explore this idea analytically and numerically. In particular
we analyze the temperature and quark mass dependence of the Polyakov
loop and heavy-quark free energy in the chiral limit.

%%%%%%%%%%%%%%%%%%%%%%%%%%%%%%%%%%%%%%%%%%%%%%%%%%%%%%%%%%%%%%%%%%%%%%%%%%%%%%
%  Polyakov loop and chiral symmetry restoration
%%%%%%%%%%%%%%%%%%%%%%%%%%%%%%%%%%%%%%%%%%%%%%%%%%%%%%%%%%%%%%%%%%%%%%%%%%%%%%
\section{The Polyakov loop and chiral symmetry restoration}\label{sec:ploop}

For lattice QCD in a Euclidean space-time volume $N_\sigma^3\times N_\tau$,
the Polyakov loop and its spatial average are given by
\begin{linenomath*}\begin{equation}
  P_{\vec{x}}\equiv\frac{1}{3}\tr \prod_\tau U_4\left(\vec{x},\tau\right),
  ~~~~~~
  P\equiv\frac{1}{N_\sigma^3}
  \sum_{\vec{x}}P_{\vec{x}},
\end{equation}\end{linenomath*}
respectively. Here $U_4(\vec{x},\tau)$ is the $\SU(3)$-valued link variable
originating at space-time point $(\vec{x},\tau)$, pointing in the
Euclidean time direction. $P$ can be related to the heavy-quark free
energy by 
\begin{linenomath*}\begin{equation}\label{eq:Fav}
  F_q(T,H)  = -T\ln \ev{P(T,H)} 
  = -\frac{T}{2} \lim_{\left|\vec{x}-\vec{y}\tinysp\right|\rightarrow \infty}
\ln \ev{P^{\phantom\dagger}_{\vec{x}} P^\dagger_{\vec{y}}}.
\end{equation}\end{linenomath*}
We have made explicit in this equation the dependence of $P$ and $F_q$ on the
temperature $T$ and symmetry-breaking parameter $H\equiv m_l/m_s$. For the
remainder of these proceedings we will not explicitly write these 
dependencies to keep the notation light.
$P$ requires a multiplicative renormalization,
\begin{linenomath*}\begin{equation}
  P = e^{-c(g^2) N_\tau}\Pbare,
\end{equation}\end{linenomath*}
i.e. the renormalized $P$ appears in eq.~\eqref{eq:Fav}.
Therefore derivatives of the free energy such as 
\begin{linenomath*}\begin{equation}
  \frac{\partial F_q/T}{\partial H} 
  = -\frac{1}{\ev{P}} \frac{\partial \ev{P}}{\partial H} 
~~~~~\text{and}~~~~~
  T_c \frac{\partial F_q/T}{\partial T} 
  = -\frac{T_c}{\ev{P}} \frac{\partial \ev{P} }{\partial T}
  \label{eq:FqHandFqT}
\end{equation}\end{linenomath*}
are independent of the renormalization, which in the continuum limit
drops out in the ratio.

\subsection{The Polyakov loop as an energy-like operator}

From the perspective of Wilson's renormalization group 
\cite{Wilson:1971bg,Wilson:1971dh}, thermodynamics in the vicinity of a 
critical point can be described by an effective Hamiltonian, which is 
defined in a multi-dimensional space of operators (observables). 
These operators may be invariant under the global symmetry that gets broken 
at the critical point or may break this symmetry explicitly. In the former 
case the operator is said to be energy-like, while in the latter case it 
is magnetization-like. In this study we are concerned with
the spontaneous breaking of the global $\SU(2)_L\times \SU(2)_R$ chiral
symmetry in (2+1)-flavor QCD, which in the continuum is expected to belong 
to the 3-$d$ $\O(4)$ universality class.  The light quark chiral condensate 
is a typical magnetization-like operator for this phase transition, i.e. 
it contributes to the Hamiltonian as a symmetry-breaking operator with 
$H\sim m_l$ as a coefficient, and in this regard $H$ parameterizes the extent 
of symmetry breaking. On the other hand the Polyakov loop is purely gluonic, 
hence it is invariant under chiral transformations of the quark fields; 
therefore $\ev{P}$ as well as the heavy-quark free energy 
$F_q$ obtained from it are energy-like observables.

In the vicinity of a phase transition, energy-like observables can be
written as the sum of a regular (analytic) part, which is a Taylor series
in the symmetry-breaking parameter $H$ and the reduced temperature 
$t\equiv(T-T_c)/T_c$, and a singular
(non-analytic) part, which is described by a universal scaling function
of the scaling variable $z=z_0 t H^{-1/\beta\delta}$.
This scaling variable depends on universal critical exponents $\beta$
and $\delta$ and is rescaled by a non-universal constant $z_0$.
The scaling behavior of an arbitrary energy-like observable in the 3-$d$ 
$\O(N)$ universality class is given in Ref.~\cite{Engels:2011km}.
In particular we can write
\begin{linenomath*}\begin{equation}
                F_q/T = A H^{(1-\alpha)/\beta\delta} f'_f(z) 
                +f_{\rm reg}(T,H),
\label{Fqcritical}
\end{equation}\end{linenomath*}
where $A$ is another non-universal constant, 
$f'_f(z)={\rm d}f_f(z)/{\rm d}z$ is the derivative of the scaling 
function $f_f(z)$ that characterizes the singular part of the logarithm of 
the partition function, $\alpha$ is another critical exponent, and 
\begin{linenomath*}\begin{equation}
                \freg= \sum_{i,j}  a^r_{i,2j}\ t^i H^{2j} 
                \equiv \sum_j p^r_{2j}(T) H^{2j}.
\label{freg}
\end{equation}\end{linenomath*}

We take eq.~\eqref{Fqcritical} as the starting point for our analysis.
For example from this and eq.~\eqref{eq:Fav} we can get $\ev{P}$ by 
\begin{linenomath*}\begin{equation}
                \ev{P} = \exp\left(-
                A H^{(1-\alpha)/\beta\delta} f'_f(z) 
                -f_{\rm reg} \right).
                \label{Pcritical}
\end{equation}\end{linenomath*}
Making use of the relation between $f_f(z)$ and the scaling function
of the order parameter $f_G(z)$,
\begin{linenomath*}\begin{equation}\label{eq:fGandff}
f_G(z) = -\left(1+\frac{1}{\delta}\right) f_f(z)
+\frac{z}{\beta\delta}f'_f(z)~,
\end{equation}\end{linenomath*}
we obtain near $T_c$
\begin{linenomath*}\begin{equation}
\frac{\partial F_q/T}{\partial H} = -A H^{(\beta -1)/\beta\delta}f'_G(z) 
             +\frac{\partial f_{\rm reg}}{\partial H} .
\label{Fqm-crit}
\end{equation}\end{linenomath*}
Furthermore, one finds the $T$-derivative of $F_q/T$ to be
\begin{linenomath*}\begin{equation}\label{eq:FqT}
T_c\frac{\partial F_q/T}{\partial T} =  
A z_0 H^{-\alpha / \beta\delta} f''_f(z) + T_c \frac{\partial f_{\rm reg}}{\partial T}  .
\end{equation}\end{linenomath*}

\subsection{Polyakov loop observables in 3-d O(2) systems}

As mentioned, the continuum limit universality class is 
expected to be $\O(4)$, but this
study works at fixed $N_\tau$ using the staggered fermion
discretization scheme, so the relevant universality
class is 3-$d$ $\O(2)$. The critical exponents, taken from Ref.
\cite{Engels:2000xw}, are 
\begin{linenomath*}\begin{equation}
\beta=0.349, ~~~~\delta=4.780, ~~~~\text{and}~~~~
\alpha=2-\beta(1+\delta)=-0.0172.
\end{equation}\end{linenomath*}

Our parametrization of the 3-$d$, $\O(2)$ scaling functions is based on
data obtained in Ref.~\cite{Engels:2000xw}. 
We used $f_G$, given in that paper in the Widom-Griffiths form,
and replaced this in the form $f_G(z)$.
We use the same ansatz,
with a Taylor series for small $z$ and the asymptotic forms for 
$z\rightarrow \pm\infty$.
After finding expansion parameters for $f_G(z)$, we can determine the 
corresponding expansion parameters for $f_f(z)$.
The only missing coefficients are, in the notation of
Ref. \cite{Engels:2011km}, $c_0^+$ and $c_0^-$, which control
the asymptotic behavior of $f_f(z)$ and are of particular interest
for the analysis of energy-like observables in the limit $H=0$,
\begin{linenomath*}\begin{equation}\label{ffasym}
                f_f(z) = |z|^{2-\alpha} 
\begin{cases}
        c_0^+ + c_1^+ z^{-2\beta\delta} &,~ z \rightarrow \infty \\
        c_0^- + c_2^- (-z)^{-\beta\delta} &,~ z \rightarrow -\infty\;.
\end{cases}
        \end{equation}\end{linenomath*}
We have calculated $c_0^+$ and $c_0^-$
using eqs. (58) and (61) of Ref. \cite{Engels:2011km}. For these we find 
$c_0^+ = 2.728(30)$ and $c_0^-=2.447(40)$.
With $A^\pm=(2-\alpha) z_0^{1-\alpha} c_0^\pm$, 
the resulting universal ratio $A^+/A^-= c_0^+/c_0^- = 1.115(30)$ agrees
well with $A^+/A^- = 1.12(5)$ calculated in Ref. \cite{Cucchieri:2002hu}.
The coefficients of the sub-leading corrections in
eq.~\eqref{ffasym} are known universal numbers, 
$c_1^+\equiv -R_\chi /2 =-0.678(2) $ \cite{Cucchieri:2002hu} 
and $c_2^-=-1$.

From eq.~\eqref{ffasym} one finds at fixed temperature and for small $H$
\begin{linenomath*}\begin{equation}
                \frac{F_q}{T} \sim 
\begin{cases}
a^-(T)+A p_s^-(T)\ H &,\ T< T_c \\
a^r_{0,0}+A a_1\ H^{(1-\alpha)/\beta\delta} &,\ T=T_c \\
a^+(T)+p^+(T)\ H^2 &,\ T> T_c
\end{cases},
\label{O4criticalmass}
\end{equation}\end{linenomath*}
with $a^\pm(T)=A a_s^\pm(T)+\freg(T,0)$ as well as $p^+(T)=A p_s^+(T)+p^r_2(T)$ 
receiving contributions from both the singular and regular terms. 
For $T\le T_c$ the 
dominant quark mass dependence arises from the singular 
term only. In particular, we have
\begin{linenomath*}\begin{eqnarray}
\label{coefficients}
a_s^\pm(T) &=& (2-\alpha)\  z_0^{1-\alpha}\ c_0^{\pm}\ t |t|^{-\alpha} \nonumber \\
p_s^-(T) &=& (2-\alpha-\beta\delta)\ (-z_0 t)^{1-\alpha- \beta\delta}  \\
p_s^+(T) &=& 2 (1-\alpha/2-\beta\delta)\ c_1^+ (z_0 t)^{1-\alpha-2 \beta\delta} \nonumber,
\end{eqnarray}\end{linenomath*}

%The $H$ dependence of $F_q/T$ is dominated by the singular contribution. 
To leading order $H$-dependent corrections to $F_q/T$ and 
$\ev{P}$ are proportional to $H$ for all $T < T_c$. 
The linear dependence on $H$ reflects the contribution of Goldstone 
modes to the $\O(N)$ scaling functions in the symmetry-broken, low-temperature 
phase. This linear dependence on $H$ is also consistent with the quark mass 
dependence of heavy-light bound states \cite{Megias:2012kb,Bazavov:2013yv},
which dominate the hadronic 
contributions to $\ev{P}$ at low $T$ as 
\begin{linenomath*}\begin{equation}
\ev{P} \approx \frac{4}{3}e^{-\Delta(m_l) /T},~~~ ~~~
 \Delta (m_l)\equiv\lim_{m_h\rightarrow \infty} (M_{hl}-m_h), 
\label{Phadron}
\end{equation}\end{linenomath*}
where $\Delta(m_l)$ is the static limit for the mass of a heavy-light 
bound state $M_{hl}$ with divergent heavy quark mass $m_h$ removed
\cite{Megias:2012kb}. 
As can be shown in heavy-quark chiral perturbation theory, 
this heavy-light binding energy depends linearly on $m_l$, 
i.e. $\Delta(m_l) -\Delta(0) \sim  m_\pi^2$  
\cite{Brambilla:2017hcq}. At low $T$ the resulting linear quark mass 
dependence of $\ev{P}$ and $F_q$, arising from the thermodynamics of a 
heavy-light hadron gas, is thus consistent with the behavior 
from the singular part of 
$\O(N)$ scaling functions valid near $T_c$. 

As $\alpha<0$, the first two terms in the regular part, 
$a_r(T)=a^r_{0,0}+a^r_{1,0} t$, dominate the 
temperature dependence and slope of $F_q/T$ and $\ev{P}$ 
at the critical point. For instance
\begin{linenomath*}\begin{equation}
  \frac{F_q(T,0)}{T} 
      = a^r_{0,0} + t \left( a^r_{1,0} + A^\pm |t|^{-\alpha} \right).
\label{FqH0}
\end{equation}\end{linenomath*}
Although at $T_c$ the contribution to the slope is entirely given by
the regular term $a_{1,0}^r$, close to $T_c$ this
contribution gets to a large extent canceled by
the singular contributions, $A^\pm |t|^{-\alpha}$.
This is the origin of the well known spike in
specific-heat like observables ($2^{\rm nd}$
derivatives of the logarithm of the partition function
 with respect to $T$) in $\O(N)$
universality classes\footnote{The appearance of this spike in
$\partial (F_q/T)/\partial T$ along with a detailed analysis
of its features is presented in Ref.~\cite{clarke_sensitivity_2020}.}.
As $|\alpha|$ is quite small, the correction from the 
singular part varies little in a large temperature range, {\it e.g.}
$|t|^{-\alpha}$ equals 0.92 for $t=0.01$ and rises to 0.96 for $t=0.1$. 
The singular part thus contributes an almost constant term to the
slope of $F_q(T,0)/T$ at $T_c$, and the change in slope that arises from the 
singular contribution is difficult to detect\footnote{Distinguishing 
singular and regular contributions by just analyzing the temperature 
dependence of $F_q(T,0)/T$ or $\ev{P}$ would require an analysis at very 
small values of $H$ in a tiny temperature interval. This, in fact, has
been done in studies of 3-$d$, $\O(2)$ spin models \cite{Engels:2000xw}
but is out of reach for current studies in QCD.}. 

In order to examine the sensitivity of $F_q/T$ and $\ev{P}$ to chiral 
symmetry restoration it thus seems easier to first analyze their
dependence on $H$ rather than $T$. Results for $\partial(F_q/T)/\partial H$
will be given in Section~\ref{sec:results}, followed by results for
$F_q/T$ and $\ev{P}$, while results for the
$H$-derivative of $\ev{P}$, the mixed susceptibility
$\chi_{mP}$, are given in Ref.~\cite{clarke_sensitivity_2020}.

%%%%%%%%%%%%%%%%%%%%%%%%%%%%%%%%%%%%%%%%%%%%%%%%%%%%%%%%%%%%%%%%%%%%%%%%%%%%%%
%  Computational setup
%%%%%%%%%%%%%%%%%%%%%%%%%%%%%%%%%%%%%%%%%%%%%%%%%%%%%%%%%%%%%%%%%%%%%%%%%%%%%%
\section{Computational setup and data analysis}\label{sec:setup}

\begin{table}
\centering
\begin{tabularx}{\linewidth}{LcC|CcR} \hline\hline
$m_s/m_l$ & $N_\sigma$ & avg. \# TU &
$m_s/m_l$ & $N_\sigma$ & avg. \# TU \\
\hline
20   &  32   & 99 000&
80   &  56   & 35 000\\

27   &  32   & 1 500 000&
80   &  40   & 33 000\\

40   &  40   & 110 000&
80   &  32   & 73 000\\

     &       &        &
160  &  56   & 17 000\\
\hline\hline
\end{tabularx}
\caption{Summary of parameters used in this study and corresponding
         statistics, reported in average molecular dynamic time units (TU)
         per parameter combination.}
\label{tab:latsnstats}
\end{table}

\begin{figure}
  \centering
  \includegraphics[width=0.67\textwidth]{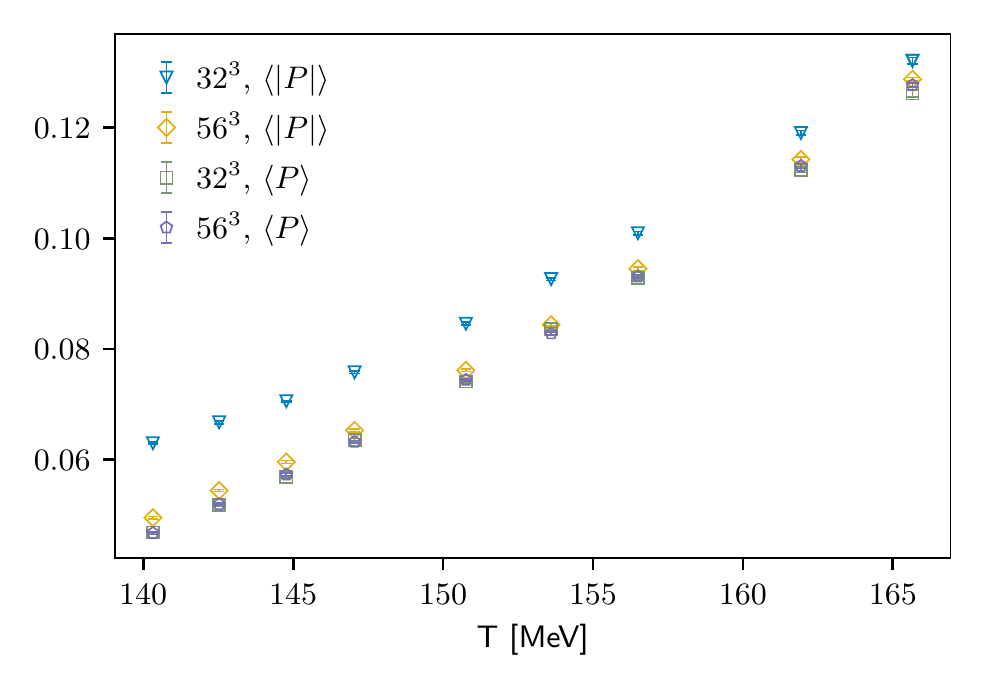}
    \caption{(Color online) Volume dependence of Polyakov loop observables for
           $N_\tau=8$, $m_s/m_l=80$ lattices for two values of the 
           spatial lattice extent $N_\sigma=32$, 56.}
   \label{fig:Pvol}
\end{figure} 

We analyze properties of (2+1)-flavor QCD where 
$m_s$ is fixed to its physical value and $m_l$ is varied in the range 
$H=m_l/m_s=1/160-1/20$.
The analysis is performed on sets of gauge field configurations
generated using highly improved staggered quarks (HISQ) 
\cite{follana_highly_2007} and the tree-level improved Symanzik gauge action. 
We utilize gauge field ensembles that were generated
previously by the HotQCD collaboration
\cite{Bazavov:2011nk,Bazavov:2014pvz,bazavov_qcd_2017,Ding:2019prx}.
Additionally, we have generated further configurations for $H=1/40$ and
$H=1/80$.

The bare coupling $\beta=6/g^2$ for physical and smaller-than-physical 
$m_l$ is taken in the range 6.26-6.50, which is chosen so 
temperatures lie in the vicinity of the chiral pseudo-critical 
temperature. For $H=1/20$ we also use data from calculations on lattices 
at smaller couplings, $\beta=6.05, 6.125$, and $6.175$ 
\cite{Bazavov:2013yv}, which allows to establish contact to the low 
temperature regime. To set the scale we use the recent parametrization 
of lattice QCD results for the kaon decay constant, $f_K\,a(\beta)$ 
\cite{Bazavov:2019www}. 
In particular, this defines our temperature scale, 
$T/f_K= 1/N_\tau f_K\,a(\beta)$, with 
$f_K=156.1/\sqrt{2}$~MeV~\cite{Bazavov:2010hj}.
Additive renormalization constants $c(g^2)$ are based on Table V of
Ref.~\cite{Bazavov:2016uvm}. We fit these constants with a univariate
spline, which is necessary to determine
renormalization constants for a few $\beta$ not listed in that table.
The results from the interpolation are taken as our $c(g^2)$.

Lattice sizes, quark masses, and statistics are summarized in
Table~\ref{tab:latsnstats}.
All calculations were performed on lattices with $N_\tau=8$. 
For $H =1/80$ we have results for aspect ratios $N_\sigma/N_\tau = 4-7$. 
We see in Fig.~\ref{fig:Pvol} no 
significant volume dependence of $\ev{P}$ in the entire
temperature range. Results on different size lattices agree within errors,
which are about 1\%. We thus can safely neglect any finite volume
corrections to our results for $\ev{P}$. This is quite different for 
$\ev{|P|}$, which shows a strong volume dependence and converges to 
$\ev{P}$ only in the infinite volume limit.

%%%%%%%%%%%%%%%%%%%%%%%%%%%%%%%%%%%%%%%%%%%%%%%%%%%%%%%%%%%%%%%%%%%%%%%%%%%%%%
%  Results
%%%%%%%%%%%%%%%%%%%%%%%%%%%%%%%%%%%%%%%%%%%%%%%%%%%%%%%%%%%%%%%%%%%%%%%%%%%%%%
\section{Results}\label{sec:results}

\begin{figure}
\centering
\hspace{-5mm}
\includegraphics[width=0.343\textwidth]{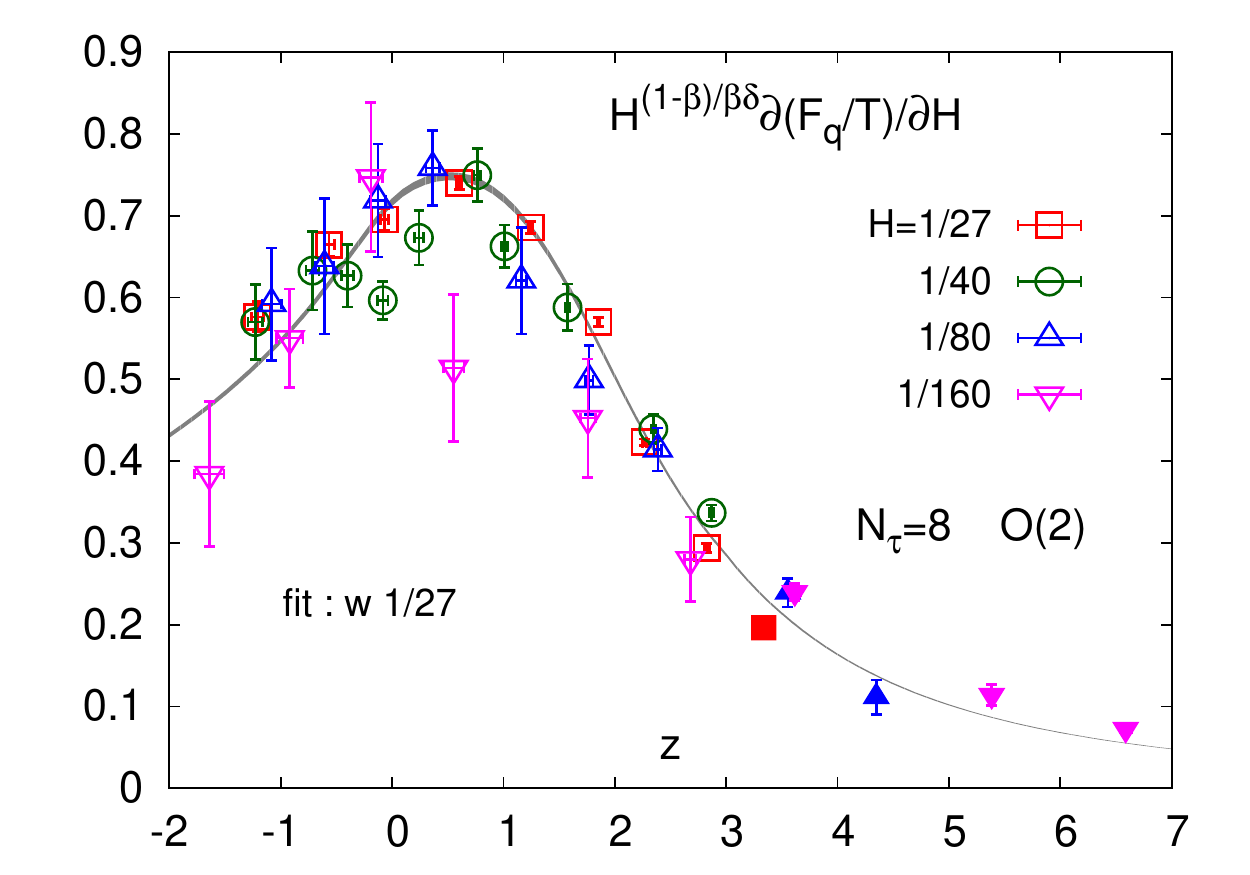}
\hspace{-3mm}
\includegraphics[width=0.343\textwidth]{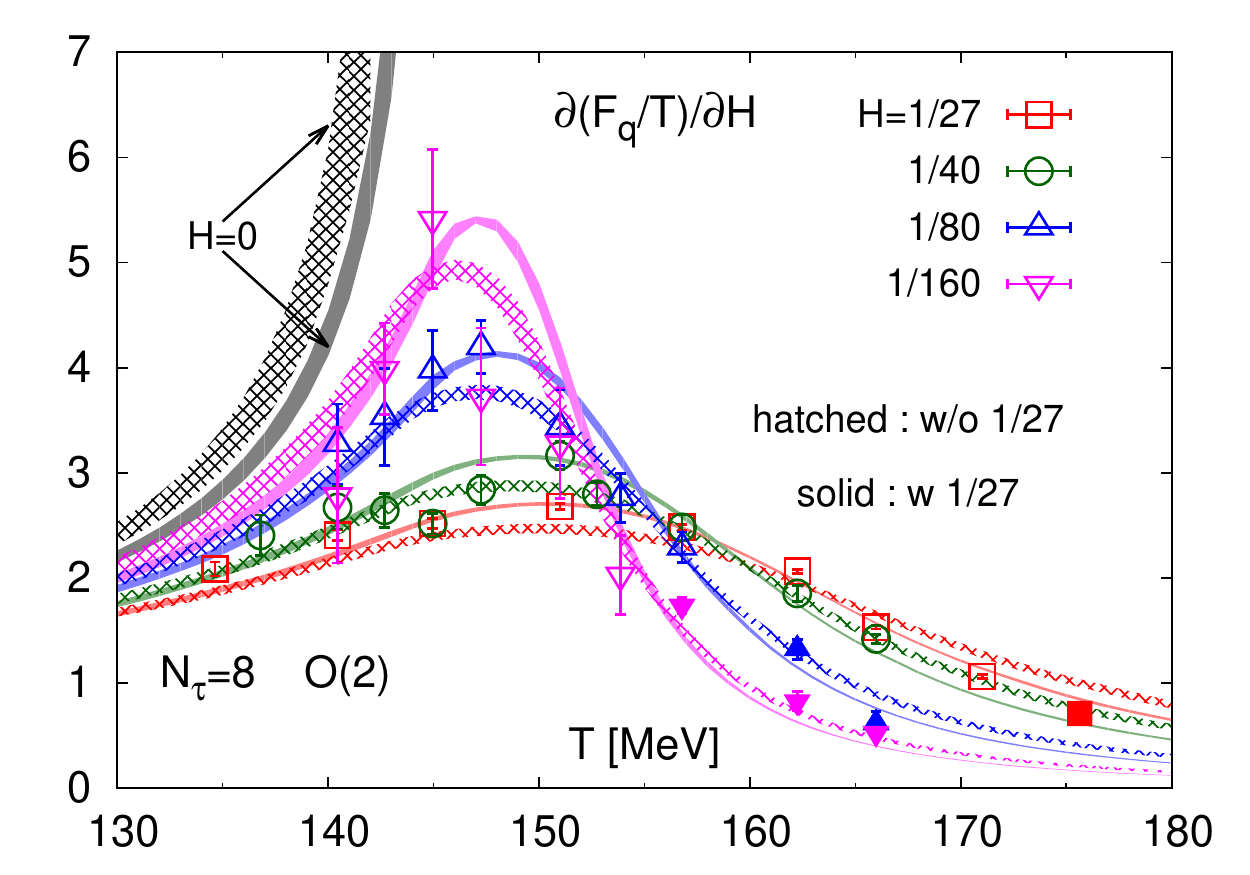}
\hspace{-3mm}
\includegraphics[width=0.343\textwidth]{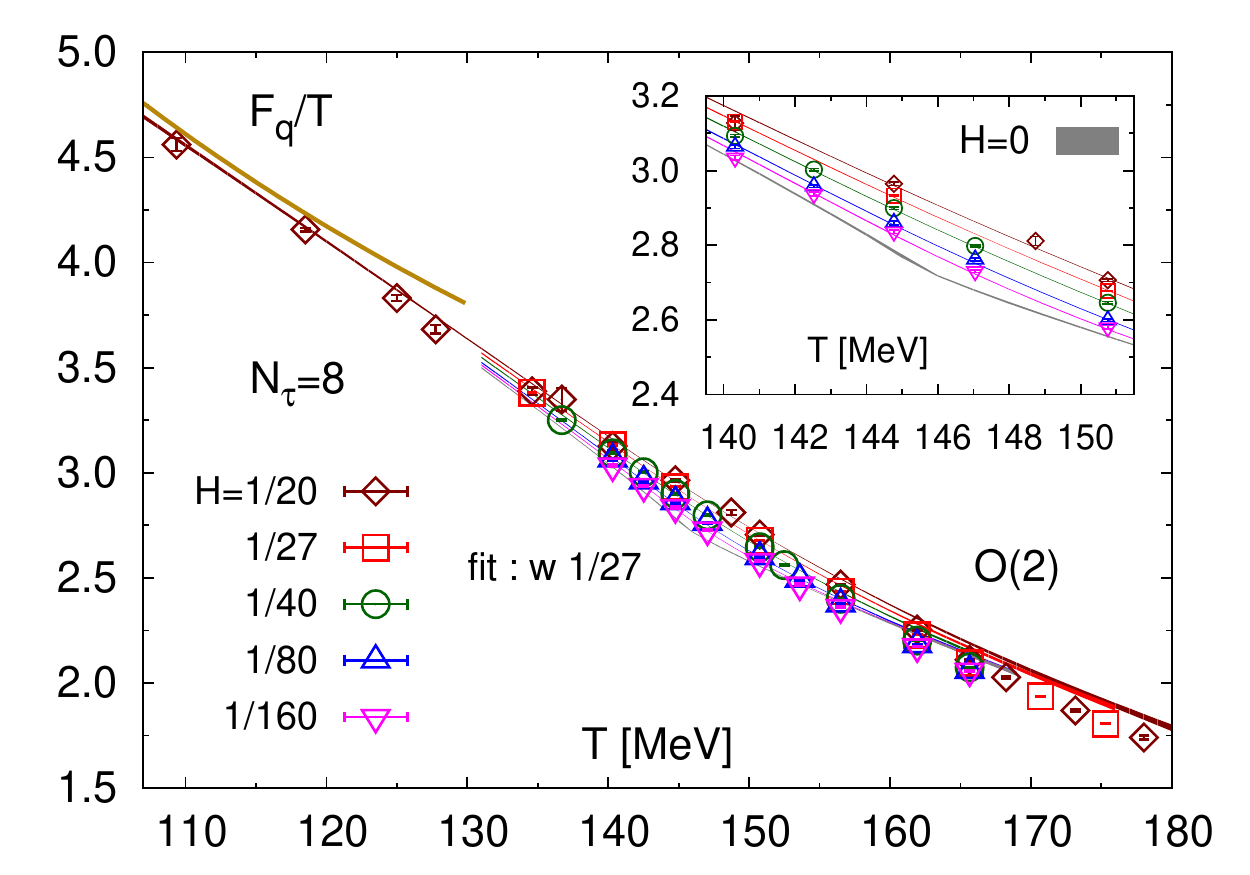}
\caption{(Color online) Scaling and temperature dependence of $F_q/T$ and 
         its $H$-derivative.
         Fits are described in the text.
         Data with filled symbols are not included in the fits.
         {\it Left}: Rescaled derivative of $F_q/T$ with respect to $H$ 
         as function of $z$. {\it Middle}: Derivative of $F_q/T$ with 
         respect to $H$ as function of $T$. {\it Right}: $F_q/T$ 
         as function of $T$. The chiral limit result 
         for $H=0$ obtained from these fits is shown as grey bands. 
         The solid gold line shows the heavy-light 
         meson contribution calculated in the hadron-gas approximation 
         \cite{Bazavov:2013yv}. Images taken from Ref.
         \cite{clarke_sensitivity_2020}.}
\label{fig:Fqfit}
\end{figure}

In Fig.~\ref{fig:Fqfit} (left) we show $\partial (F_q/T)/\partial H$
rescaled with the appropriate power of $H$ expected
from the $\O(2)$ scaling ansatz. The good scaling behavior
suggests that $\partial (F_q/T)/\partial H$ will indeed diverge 
as $H^{(\beta-1)/\beta\delta}$ in the chiral limit and that $H$-dependent
contributions to $F_q/T$ arising from regular terms are small compared to
those coming from the singular part. This motivates an
$H$-independent fit ansatz for $\freg$, 
i.e. we use $\freg (T,H=0)$ in all our fits.

A fit to $\partial (F_q/T)/\partial H$ thus only involves
the singular term of eq.~\eqref{Fqm-crit}. We performed this
3-parameter fit for all data sets by either 
including or leaving out the data for
$H=1/27$. These fits are shown in
Fig.~\ref{fig:Fqfit}~(middle); the corresponding fit parameters,
$A$, $T_c$, $z_0$, are given in Ref. \cite{clarke_sensitivity_2020}. 
In particular, we find that $T_c$ and $z_0$ obtained from these fits
agree well with earlier results for chiral susceptibilities
in (2+1)-flavor QCD \cite{Ding:2019prx}.

Data for $F_q/T$ are shown in Fig.~\ref{fig:Fqfit} (right).
They have been fitted to eq.~\eqref{Fqcritical} using only the 
constant and linear $H$-independent
terms in the regular part as fit parameters and keeping fixed
the three non-universal constants, determined in the 
previous step, in the singular part. The $T$-range and data 
used in the fit are shown in the inset. The  resulting fit 
parameters $a^r_{0,0}$ and $a^r_{1,0}$ are also given in Ref.
\cite{clarke_sensitivity_2020}. In this figure we also show the 
heavy-light meson contribution to $F_q/T$ calculated in the hadron-gas 
approximation \cite{Megias:2012kb,Bazavov:2013yv}. 
As mentioned earlier, chiral perturbation theory also gives a 
linear dependence on $H$ at 
low temperature \cite{Brambilla:2017hcq}.

\begin{figure}
\centering
\includegraphics[width=0.495\textwidth]{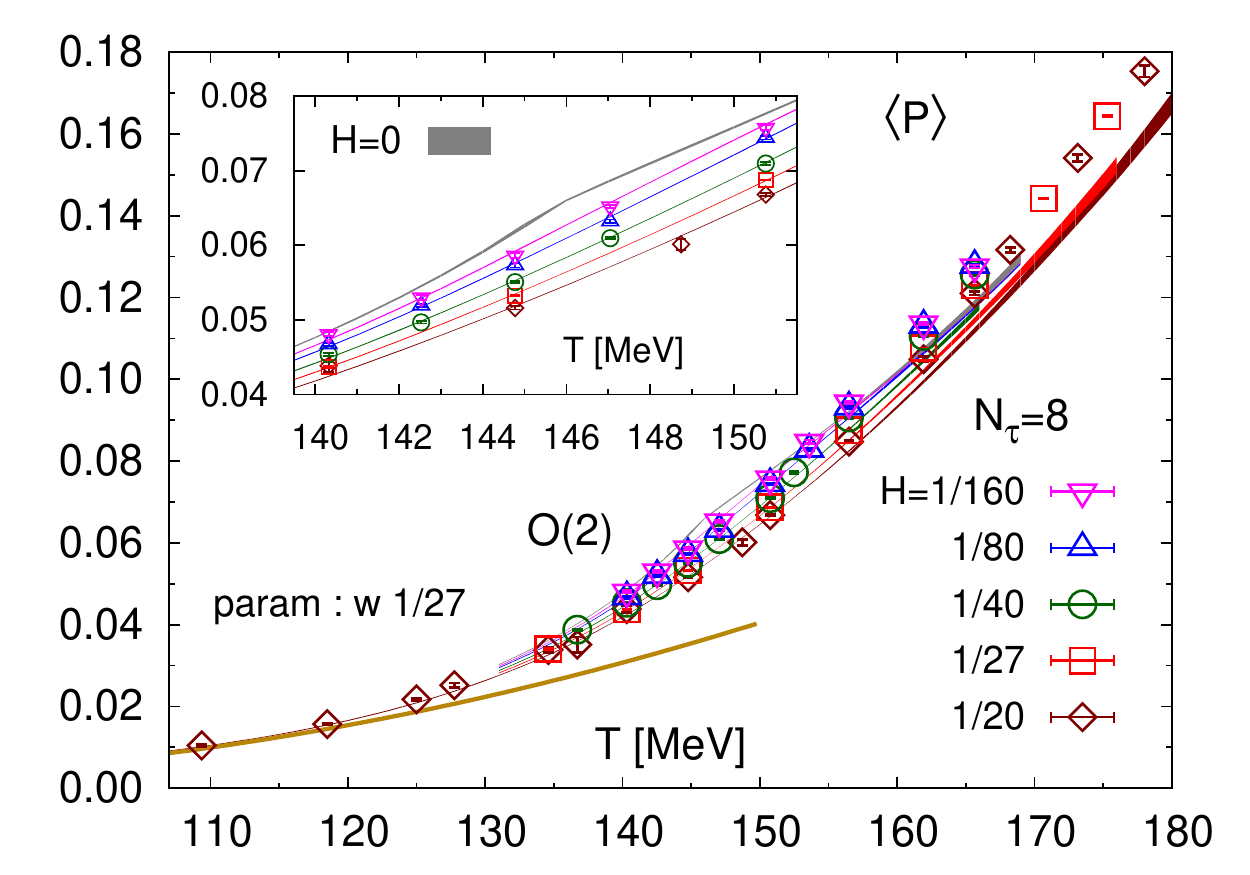}
\includegraphics[width=0.495\textwidth]{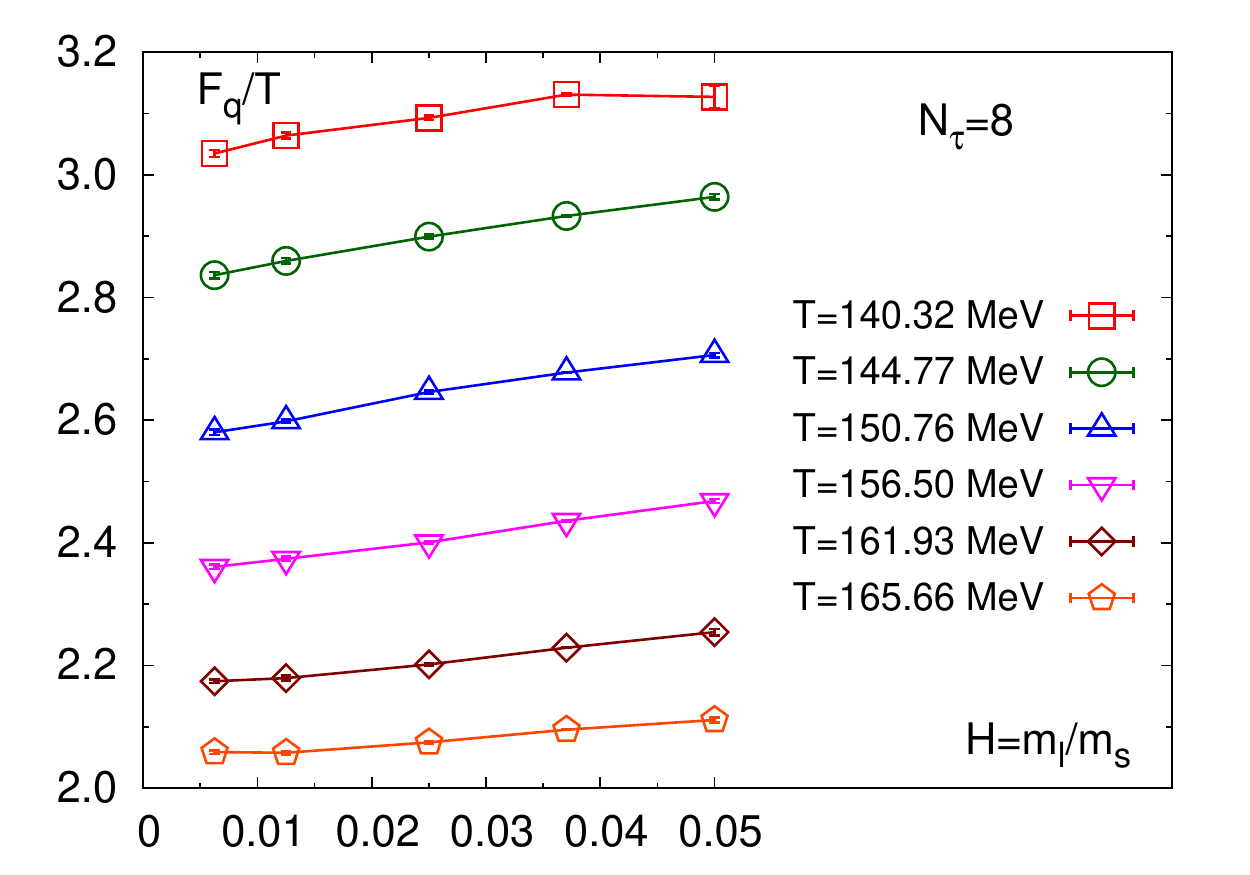}
  \caption{(Color online) 
           {\it Left:} $T$-dependence of $\ev{P}$ for several values of $H$.
           The inset shows data in the $T$ range covered by the fit. 
           The chiral limit result for $H=0$ obtained
           from this fit is shown as a grey band. The solid gold 
	   line is as described in Fig.~\ref{fig:Fqfit}.
           Image taken from Ref. \cite{clarke_sensitivity_2020}.
           {\it Right:} $H$-dependence of $F_q/T$ for several values of $T$,
           with lines to guide the eye.}
            
  \label{fig:PmassTall}
\end{figure}

Once we have determined all five fit parameters for $F_q/T$, we can 
plug them into eq.~\eqref{Pcritical} to arrive at a parameter-free 
description of the $T$ and $H$ dependence of $\ev{P}$. The thus 
determined curves are shown in Fig.~\ref{fig:PmassTall} (left). 
As seen in the inset, they agree well with $\ev{P}$
data near $T_c^{N_\tau=8}=144(2)$~MeV \cite{Ding:2019prx}, 
which suggests the behavior of $\ev{P}$ is explained
well by chiral scaling in this region and serves
as a consistency check of our approach.

Turning back to Fig.~\ref{fig:Fqfit} (middle),
it is apparent that $\partial (F_q/T)/\partial H$
decreases with $H$ at high $T$ and increases at low $T$.
This is consistent with an approach towards zero at high 
$T$ and a non-vanishing, strongly temperature-dependent constant
at low $T$ in the chiral limit. 
Such a pattern is in accordance with the expected quadratic
dependence on $H$ for $T>T_c$ and the leading linear dependence for $T<T_c$
given in eq.~\eqref{O4criticalmass}. For $T<T_c$ the next-to-leading
dependence, $H^{3/2}$, will come with a negative coefficient; 
correspondingly there is a change from a convex (at high $T$) to concave
(at low $T$) $H$-dependence evaluated at fixed temperature 
in Fig.~\ref{fig:PmassTall}~(right).
Although errors are large for the results obtained with
the smallest light quark mass, $H=1/160$, it is evident that 
$\partial (F_q/T)/\partial H$ has maxima at
$T\sim 145-150$~MeV which are close to $T_c^{N_\tau=8}$. 
With decreasing $H$ they
approach $T_c^{N_\tau=8}$, and the peak height increases.

%%%%%%%%%%%%%%%%%%%%%%%%%%%%%%%%%%%%%%%%%%%%%%%%%%%%%%%%%%%%%%%%%%%%%%%%%%%%%%
%  Conclusions and Acknowledgements
%%%%%%%%%%%%%%%%%%%%%%%%%%%%%%%%%%%%%%%%%%%%%%%%%%%%%%%%%%%%%%%%%%%%%%%%%%%%%%
\section{Conclusions}\label{sec:conclusions}

We examined the $m_l$ dependence of $\ev{P}$ and $F_q$. 
With this work we did not aim at presenting continuum-extrapolated 
results but rather wanted to establish evidence for the influence of 
chiral symmetry restoration.
The $m_l$ derivative of $F_q$ diverges in the chiral limit consistent 
with the expected behavior for energy-like observables in 
the 3-$d$, $\O(2)$ universality class. The data for $\ev{P}$
are described well by $\O(2)$ scaling near $T_c$. 

Finding evidence for critical behavior in the temperature derivatives of 
$F_q$ and $\ev{P}$ is challenging. A clear singular
structure will only build up at very small values of the quark masses. 

As the critical exponent $\alpha$ is also negative and small in the 
$\O(4)$ universality class, we expect similar behavior for 
$F_q(T,H)/T$ and $\ev{P}$ will also persist in the 
continuum limit.

\section*{Acknowledgements}
This work was supported by the Deutsche Forschungsgemeinschaft
(DFG, German Research Foundation) Proj. No. 315477589-TRR 211; and by
the German Bundesministerium f\"ur Bildung und Forschung through
Grant No. 05P18PBCA1.
We thank HotQCD for providing access to their latest data
sets and for many fruitful discussions.

%%%%%%%%%%%%%%%%%%%%%%%%%%%%%%%%%%%%%%%%%%%%%%%%%%%%%%%%%%%%%%%%%%%%%%%%%%%%%%
%  References 
%%%%%%%%%%%%%%%%%%%%%%%%%%%%%%%%%%%%%%%%%%%%%%%%%%%%%%%%%%%%%%%%%%%%%%%%%%%%%%
\printbibliography

\end{document}